\documentclass[a4paper]{article}
\pdfoutput=1
\usepackage[english]{babel}
\usepackage[utf8x]{inputenc}
\usepackage[T1]{fontenc}
\usepackage[a4paper,top=3cm,bottom=2cm,left=3cm,right=3cm,marginparwidth=1.75cm]{geometry}
\usepackage{changes} % use \usepackage[final]{changes} for clean version

\usepackage[section]{placeins}

\usepackage{amsmath}
\usepackage{graphicx}
\usepackage[colorinlistoftodos]{todonotes}
\usepackage[colorlinks=true, allcolors=blue]{hyperref}

\title{Performance and sensitivities of home detection from mobile phone data}
\date{} %If commented out, you get the date of today automatically

\author{Maarten Vanhoof, Clement Lee, Zbigniew Smoreda}

\begin{document}
\maketitle

\begin{abstract}
Large-scale location based traces, such as mobile phone data, have been identified as a promising data source to complement or even enrich official statistics. In many cases, a prerequisite step to deploy the massively gathered data is the detection of home location from individual users. The problem is that little research exists on the validation (comparison with ground truth datasets) or the uncertainty estimation of home detection methods, not at individual user level, nor at nation-wide levels. In this paper, we present an extensive empirical analysis of home detection methods when performed on a nation-wide mobile phone dataset from France. We analyze the validity of 9 different Home Detection Algorithms (HDAs), and we assess different sources of uncertainty. Based on 225 different set-ups for the home detection of around 18 million users we discuss different measures for validation and investigate sensitivity to user choices such as HDA parameter choice and observation period restriction. Our findings show that nation-wide performance of home detection is moderate at best, with correlations to ground truth maximizing at 0.60 only. Additionally, we show that time and duration of observation have a clear effect on performance, and that the effect of HDA criteria and parameter choice are rather small compared to other uncertainties. Our findings and discussion offer welcoming insights to other practitioners who want to apply home detection on similar datasets, or who are in need of an assessment of the challenges and uncertainties related to mobilizing mobile phone data for official statistics.
\end{abstract}

\section{Introduction}
\subsection{Mobile phone data and official statistics}
Recently, there has been a growing interest in the use of ‘Big Data’ in official statistics \cite{Daas2015,Vanhoof_JOS}. One reason is because ‘Big Data’ sources can offer more timely, high-resolution information compared to standard statistics production. Another reason is that digital data sources, such as mobile phone data, can now guarantee reproducibility in different countries, at different spatial scales, and for different times thanks to the ubiquitous use of mobile phones by large populations \cite{Pappalardo2016}. 

As a consequence, numerous recent works have pointed out the potential of mobile phone datasets to complement existing official statistics. Call Detailed Record (CDR) data, the most basic type of mobile phone data collecting information at every call or text of an individual user, for example, have shown to allow for an estimation of population presence on a national scale and for any given time period, forming an interesting complement to population and migration statistics \cite{Deville2014}. Furthermore, CDR data have been used to estimate general commuting times and distances \cite{Kung2014}, to estimate directions of general mobility of populations during very short timeframes \cite{Balzotti2018} or to reproduce the French official urban area classification \cite{Vanhoof_miningurbanareas}. At a more individual level, the movement patterns captured in CDR data offer opportunities to extract figures on domestic tourism trips \cite{Vanhoof_domestictourism}, long-distance travel \cite{Janzen2018_closer}, or even to impute purpose of long-distance travel (\cite{Janzen_Purpose}), all of which offer great complementary sources to expensive travel surveys.

Other lines of research have revealed how nation-wide indicators on human mobility, calling activities and subscription purchases derived from mobile phone data relate directly to measures of deprivation, poverty, or even food security as provided by official statistics \cite{Pappalardo2016,Eagle2010,Frias-martinez2013,Decuyper2014,Vanhoof_entropy}. Following such observations, researcher have suggested that mobile phone data could be used to \textit{nowcast} or even predict socio-economic indicators at nation-wide level and for a rather high temporal and spatial resolution, marking their potential for integration in official statistics \cite{Pappalardo2015,Giannotti2012,Pappalardo2016}. 

In other words, CDR data have manifested themselves as an interesting, seemingly trustworthy addition to current official statistics. Still, incorporation of CDR data in official statistics has not (yet) been carried out at large, despite substantial investments of governments and institutions alike. One of the reasons is that big data sources, and in extension their related methodologies, often do not adhere to the standards and principals of official statistics, such as coverage, representativeness, quality, accuracy, and precision \cite{Daas2015,Vanhoof_JOS}. As such, the integration of Big Data in official statistics demands a certain investigation and renegotiation of principles or, in this case, even an organizational reinvention of official statistics offices. The importance of such investigation and renegotiation, however, is not to be underestimated as it forms an essential step in safeguarding the quality and transparency of official statistics and, in extension, of big data practices. This is importance because, due to the high speed in which big data applications are currently developed, attention to quality assessment, sensitivity, or representativeness of big data practices is not always guaranteed, even in academic work.

\subsection{The home detection problem}

One specific example of underdeveloped quality assessment for big data sources is the example of home detection methods for geo-located traces that are captured on a non-continuous base, such as mobile phone data or check-in data on location based social networks (such as Flickr, TripAdvisor or Foursquare). Many of the works that use such geo-located traces, including most of the aforementioned works using CDRs, have home detection as a prerequisite step for their analysis. Specifically for CDRs, detecting a home location for each individual user means that one cell tower - the one that is most likely to cover the actual living place of an user - is identified as \textit{home}. Whether a cell tower is more or less likely to cover the actual living place of an user is determined by so-called Home Detection Algorithms (HDA) which are deployed on a historical record of geo-locations produced by the user. 

Detecting home locations for individual users is important in different ways for different strands of mobile phone data research. To infer long-distance travel, tourism trips, or commuting, for example, home detection is needed in order to determine when a user is performing a specific type of mobility or not. Regarding the relation between mobile phone indicators and socio-economic indicators, home detection is needed too. Because socio-economic indicators are mostly available at aggregated levels, mobile phone indicators need to be aggregated in space to become comparable with, for example, census data. These aggregations, are typically done using home detection, meaning mobile phone users get aggregated based on their assumed home locations. 

Although home detections by means of HDAs are widely deployed in mobile phone data research, little discussion exists on their validity. As noted by \cite{Vanhoof_JOS}, validation of home detection is generally carried out by comparing aggregated user counts with population counts available from census data. But such high-level validation does not allow performance assessment at the individual level, which obscures insights on quality (how correct is home detection for a specific user), representativeness (how correct is home detection for different subgroups of users), or sensitivity to researchers’ decisions such as algorithm choice, criteria choice, parameter choice, duration of observation, and period of observation (how do these choices influence quality and representativity issues). 

\subsection{Exploring sensitivities of home detection}
For two main reasons it is understandable that in-depth investigations on home detection methods are typically not part of studies that use them as a prerequisite step. First, proper validation data is not always easy to obtain. In many African countries, for example, even high-level ground truth data in the from of population counts might be absent or inaccessible. Individual validation data is even harder to obtain. Although operators might possess customer related information such as billing addresses, in many countries, due to privacy regulations, it is not allowed to pair this data with mobile phone datasets. Secondly, it requires multiple iterations to investigate the performance of home detection. With the size of current CDR datasets iterations are technically challenging, time consuming and computationally expensive, even with current big data technologies, because of the many look-up functions that need done in order to collect observations from all individual users.   

Still it remains inexplicable that, apart from \cite{Vanhoof_JOS,Bojic2015}, no studies have focused on the performance of home detection methods and their sensitivity to, for example, HDA choice or the duration of available data. Concerning HDA choice, \cite{Vanhoof_JOS,Bojic2015} found that the percentage of users that get attributed a different home location when a different HDA was deployed range from 1\% to 9\% when using data on credit card transactions in Spain, from 7\% to 20\% when using a worldwide Flickr data and from 4\% to 40\% when using a CDR dataset in France (the same one as we will deploy). HDA choice, in other words, may significantly influence home detection at nation-wide scale. As a consequence, it becomes primordial to investigate the validity and sensitivities of current home detection methods, especially when such methods are considered for official statistics.   

In the case of home detection methods, there are several consequences to the absence of studies that assess validity, sensitivity, or that perform error estimation. The direct consequence is that it currently remains unclear what is influencing the quality of home detection methods. The broader consequence is that many assumptions underlying home detection methods remain unproven or even uncovered. One assumption, for example, is that the period and duration of observations do not influence the quality of home detection. Common sense suggest that this is not the case, as does research. \cite{Deville2014,Vanhoof_domestictourism} for example show how during summer months mobile phone users in France tend to move to touristic areas along the coast or near the mountains. Performing home detection on mobile phone data collected during this period will more likely lead to more wrongful results compared to using another period, proving one underlying assumption wrong.

The goal of this paper is to empirically explore the nation-wide performance of home detection methods with a focus on the sensitivity to user choices such as the HDA-choice, the chosen period of observation and the chosen duration of observation and for a case study in France. A clearer insight into the combined effects of user choices on the quality of home detection is desirable, and can help future work when making user choice as well as it can inform on the uncertainty and error related to home detection methods when performed on CDR data.

\section{Deploying home detection algorithms to a French CDR dataset}
\subsection{The French mobile phone dataset}

In our analysis, we will use an anonymized mobile phone dataset recorded by the French Operator Orange. The dataset covers mobile phone usage of ~18 million subscribers on the Orange network in France during a period of 154 consecutive days in 2007 (May 13, 2007 to October 14, 2007). Mobile phone penetration is being estimated at 86\% at that time and given a population of 63.945 million inhabitants during the observed period, that is about 32\% of all French mobile phone users and 28\% of the total population.

The mobile phone dataset consists of Call Detailed Record (CDR) data, which are typically collected by mobile phone service providers for billing and network maintenance purposes. Every time a call or text is initiated or received, CDR data store locational (the used cell tower)\added{,} temporal (time and duration of usage)\added{,} and interactional (who contacts whom) information for both correspondents. Location traces from CDR data thus are non-continuous as they are user initiated and rather sparse in time. In compliance with ethical and privacy guidelines CDR data are anonymized.

The presented dataset is one of the largest CDR datasets available worldwide in terms of population-wide coverage and has been extensively studied (e.g. \cite{Sobolevsky2013_delineating,Deville2014,Grauwin2017,Janzen2018_closer,Vanhoof_JOS,vanhoof_arxiv_spatial_uncertainty}). It is the latest CDR dataset available for France that allows for such a long term, continuous temporal —but anonymised— tracking of users in France. More recent datasets are aggregated and/or re-anonymized every given time-period in rule with The French Data Protection Agency (CNIL), who is anticipating the EU General Data Protection Regulation that does not allow collecting individual traces extensively as they are considered risky even if personal identification information is irreversibly recoded.

The spatial accuracy of the dataset is restricted to the spatial resolution of the network, that is, to the locations of the cell towers installed by the network provider. The spatial distribution of the 18,273 cell-tower locations is known but not uniform. In general, higher densities of cell towers are found in more densely populated areas like cities or coastlines. Lower densities of cell towers are observed in more rural areas, as well as in mountain or natural reserve areas. The Voronoi tessellation of all cell tower locations is shown in figure \ref{fig:map_logratio} illustrating the coverage and the density of the network.

The temporal resolution of the analyzed dataset is inhomogeneous as CDR are only created and stored during calls thereby generating records on both caller and callee side. For example, for one arbitrary day of the covered timespan (Thursday, 1st October 2007), the median number of records per user was four, relating to only two different locations. Such statistics are representative for CDR based studies and can be deemed rather high compared to other large-scale non-continuous datasets like credit-card transactions or Flickr photos \cite{Bojic2015}.

On one hand, temporal sparsity in observations and spatially inhomogeneous distributions of covered areas are typical characteristics of CDR datasets and pose substantial challenges for their automated analysis and the quality of home detection. On the other hand, the very large scale reach at population level without requiring active participation of the user for location sharing whilst at the same time preserving anonymity as well as privacy is very attractive for many application areas. Aggregating data over extended periods of time enables complex analysis and diminishes influence from singular events and/or non-routine behavior.

\subsection{Defining nine simple home detection algorithms}

Most HDAs that are deployed on CDR data consist of single-step approaches, which detect a \textit{home} by selecting the cell tower that accords best to an imposed decision rule. This is opposed to two-step approaches where spatial grouping of cell towers is performed as an extra step. The decision rules applied in HDAs can be simple or complex, meaning that they are based on one criterion or several criteria, respectively \cite{Vanhoof_JOS}. We opt to use simple decision rules over complex decision rules, as this allows better singling out the effect of criteria choice. Examples of typical criteria are ‘home is the location where most calls are made’ and ‘home is the location where most activity has been observed during nighttime’. Clearly, some of these criteria are subject to a parameter choice. For example, in the case of the ‘nighttime’ criteria, a definition of nighttime has to be specified as, for example, between 21.00 and 07.00 hours.  

%\subsubsection{Investigating criteria choice}
Based on \cite{Vanhoof_JOS,Vanhoof_homdet_arxiv, Bojic2015}, we use three criteria to perform home detection. Note that \cite{Vanhoof_JOS} elaborate on an extra criterion related to spatial grouping but because of a consistently lower performance with regard to the other criteria, we omit this criterion in our analysis. 

\begin{itemize}
%List with bolletjes, of andere dinges als je wilt: \item[-]
\item The maximum amount of activities criterion (MA): ‘home is the cell tower where most activities of the user occurred during a specific observation period’
\item The distinct days criterion (DD): ‘home is the cell tower where the maximum active days of a user were observed during a specific observation period.’
\item The time constraints criterion (TC): ‘home is the cell tower the most activities of the user occurred between XX.xx and YY.yy hours during a specific observation period. 
\end{itemize}

%\subsubsection{Investigating parameter choice}
The advantage of the MA and DD criteria is that they do not require any parameter choice. The TC criterion on the other hand demands a parameter choice on which time restriction to deploy. Despite this extra parameter choice, home detection algorithms based on the TC criterion are rather popular, especially in the form of restricting periods to nighttime and/or weekend days. The reasons, most probably, are because this restriction is intuitive and parameter choices for how to define nighttime can sometimes be based on available time surveys, both of which lend some justification to parameter and criteria choice. Nevertheless, studies investigating the sensitivity of the TC criterion to its parameter choices are nonexistent, and neither is it clear whether nighttime or weekend days in itself are the best option to consider. For this reason, we will investigate different parameters choices for the TC criterion, incorporating either nighttimes, daytimes, week days, weekend days or a combination of them. The parameters we consider are listed in table \ref{table:hdas}. 

%\begin{itemize}
%List with bolletjes, of andere dinges als je wilt: \item[-]
%\item 19.00 to 09.00
%\item 19.00 to 09.00 and including weekend days
%\item 21.00 to 07.00
%\item 21.00 to 07.00 and including weekend days 
%\item 09.00 to 21.00
%\item 09.00 to 21.00, excluding weekend days
%\item weekend days only (Saturday 00.01 to Sunday 23.59) 
%\end{itemize}

%\subsubsection{Nine different home detection algorithms}
Nine HDAs are defined by combining different criteria and parameter choices aforementioned. They are described in Table \ref{table:hdas} and will be deployed in our analysis. Next, we specify how we will apply them to different time periods with different durations, in order to assess the influence on performance.

\begin{table}[htpb!]
\caption{Description of deployed HDAs}% title of Table
\centering % used for centering table
\resizebox{\columnwidth}{!}{%
\begin{tabular}{|p{3cm} | p{2cm} | p{2.1cm}| p{7cm}|}% centered columns (2 columns) p{5cm} defines the width of, in this case, second column
\hline %insert horizontal line
\textbf{Criteria} & \textbf{Parameters} & \textbf{Name} & \textbf{Description: 'home is cell tower where:'}\\[0.5ex] % inserts table heading, puts a 0.5 vertical whitespace under the text
\hline % inserts single horizontal line
% inserting body of the table
Maximum Amount & / & MA & Most activities occurred \\ 
Distinct days & / & DD & the maximum active days were observed  \\ 
Time constraints & 19,9 & TC-19-9 & Most activities occurred between 19pm and 9am (night-time) \\ 
Time constraints & 19,9,weekend & TC-19-9-WE & ... between 19pm and 9am (night-time) and during weekend days \\ 
Time constraints & 21,7 & TC-21-7 & ... between 21pm and 7am (night-time) \\ 
Time constraints & 21,7,weekend & TC-21-7-WE & ... between 19pm and 9am (night-time) and during weekend days \\ 
Time constraints & 9,19 & TC-9-19 & ... between 9am and 19pm (daytime) \\ 
Time constraints & 9,19,week & TC-9-19-WK & ... between 9am and 19pm (daytime) but only during weekdays \\ 
Time constraints & weekend & TC-WE & ... during weekend days only (Sat and Sun) \\ 
[1ex]% [1ex] adds vertical space
\hline%inserts single line
\end{tabular}
}
\label{table:hdas} % is used to refer this table in the text
\end{table}

\subsection{Different observation periods}
While one could expect the performance of HDAs to be dependent on the period of observation, no studies have explored this effect. We will investigate such effect by analyzing the results of the nine HDAs when deployed to CDR data that was collected during different time periods that start at different dates and last different durations. 

%\subsubsection{Duration of observation periods}
The French dataset we use comprises data from the May 13 till October 15, 2007. The most straightforward duration to consider would thus be 154 days, or the entire dataset. The disadvantages of doing so, however, are that calculations are computationally expensive, and that it does not allow sensitivity to time period to be studied. Also it is not yet proven that the longest duration necessarily leads to the best performance of home detection methods. We therefore also investigate discrete months as time period as was also done in \cite{Vanhoof_JOS}. As using discrete months obscures proper comparison because of different numbers of days in different months, we also investigate durations of exactly 14 and 30 days, representing the amount of two weeks and about one month of data. 

%The different duration of observation we will consider are thus: 
%\begin{itemize}
%List with bolletjes, of andere dinges als je wilt: \item[-]
%\item Discrete months: May (19 days), June (30 days), July (31 days), August (31 days), September (30 days), October (15 days)
%\item 154 days: Entire time period
%\item 30 days: Moving window of 30 days
%\item 14 days: Moving window of 14 days
%\end{itemize}

\begin{figure}[htpb!]
\centering
\caption{Different durations and related time periods used in the analysis.}
\label{fig:time_periods}
\resizebox{0.8\textwidth}{!}{
\includegraphics{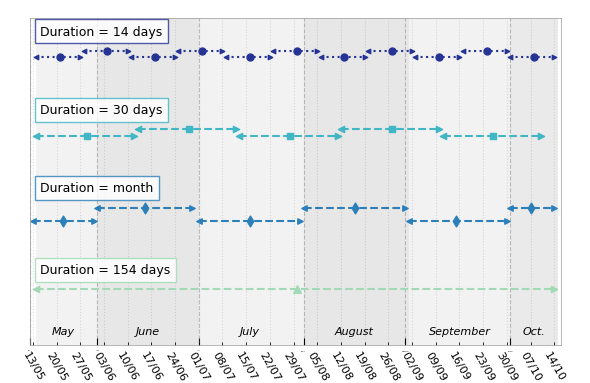}
}
\end{figure}

%\subsubsection{Time periods}
The actual time periods that relate to the different durations are straightforward for the discrete months and the 154 days duration, but not necessarily for the 14 and 30 days. For the latter, we start observations at the first day of the dataset (May 13) and take a moving window of respectively 14 and 30 days from that point in time. For the 14-day duration, this means that home detection is performed on 11 two-weeks periods, the first one being the period from May 13 till May 26, the second one being May 27 till June 9, and so on. The same strategy is used for the 30-day duration, resulting in 5 periods, the first one from May 13 till June 11. The last 8 days from October 7 till October 15 are omitted. Note that, because of our dataset starting from mid-May, the 30-day periods are more or less complementary to the discrete months, offering interesting opportunities for comparison. All time periods for all different durations used in the analysis are illustrated in figure \ref{fig:time_periods}.

\subsection{Summary of data and set-up}
Concluding, our research set-up performs home detection for each user in a French CDR dataset (around 18 million), based on nine HDAs (3 different criteria, 7 different parameter choices for the TC criterion), and for each of the twenty-three time periods (6 discrete months periods, 1 entire observation period, 5 30-day periods and 11 14-day period). This totals to around 3.7 billion home detections performed based on user location traces. Before using the home location results to assess the influence of different user decisions (criteria choice, parameter choice, time period) on performance in Section \ref{sect.results}, we first discuss how exactly we will measure performance in the next section.

\section{Assessing home detection performance at nation-wide scale}

Validation of HDAs at the individual level is not straightforward because collecting individual level ground truth data is extremely expensive and comes with increased privacy risks. As a consequence, researchers have to settle with high-level validation practices. In our analysis, we will use aggregated population counts from census data as a ground truth dataset to compare against aggregated user counts that are the results of deploying previously defined HDAs. How the ground truth dataset was constructed and which measures for comparison were used are elaborated in this section.  

\subsection{Ground truth data}

Similar to \cite{Vanhoof_JOS}, we use a ground truth dataset, which consists of population counts aggregated at the cell tower level, prepared by the French official statistics office INSEE. To construct this dataset, INSEE aggregated geo-located information of the home addresses of the French population onto the Voronoi polygons of the Orange cell tower network, which form an estimation of the coverage of each cell tower in the CDR dataset. The advantage is that no translations need to be made from census grid to the cell tower network grid (or vice versa), thus avoiding a translation error. The disadvantage, in our case, is that geo-located information on home addresses was available only for the year 2010, three years later than the mobile phone dataset was collected. This means that any comparison between ground truth population counts and user counts from CDR data has to be relative because the Orange users represent only a share of the French population. As a consequence, the assumption we introduce is that, during a 3 year period, the spatial distribution of the French population over the cell tower network does not change drastically.

\subsection{Assessing performance and sensitivities}
To measure the performance of the HDAs, we compare the outcome of each algorithm with the ground truth data. Specifically, we evaluate the degree of similarity between a vector of user counts (based on a HDA), denoted by $\vec{x}$, and a vector of population counts (based on census data), denoted by $\vec{y}$, both aggregated at cell tower level. Both vectors $\vec{x}$ and $\vec{y}$ thus have an equal length representing the 18,273 cell towers in the Orange network.

\begin{figure}[htbp!]
\centering
\caption{Overview of the different methodological steps.}
\label{fig:overzicht}
\resizebox{0.8\textwidth}{!}{
\includegraphics{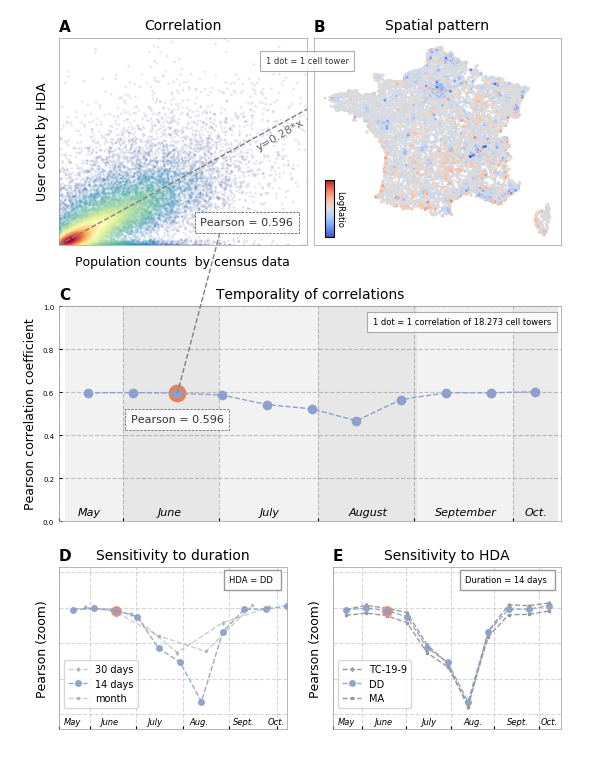}
}
\end{figure}

Because of the unknown spatial distribution of the 28\% sample of Orange users, our assessment of similarity cannot be absolute. Therefore, we define performance measures based on the relative similarities between both vectors. Once a performance measure is calculated for all nine HDAs during all time periods, we can evaluate the influence of criteria choice, parameter choice, duration of observation and period of observation on performance. The different parts of our methodology are illustrated in figure \ref{fig:overzicht}.Each sub-figure will be explained in detail in Sect \ref{sect.results}.

\subsubsection{Correlation with ground truth data (figure \ref{fig:overzicht} A) }\label{sect.pearson_definition}

One direct way of quantifying (dis)similarities is to calculate the Pearson correlation coefficient (Pearson's R) between vectors $\vec{x}$ (user counts) and $\vec{y}$ (population validation count):

\begin{equation}
Pearson's\:R (\vec{x},\vec{y})=\frac{\Sigma(x_i - \bar{x})(y_i - \bar{y})}{\sqrt{\Sigma(x_i - \bar{x})^2\Sigma(y_i - \bar{y})^2}}
\end{equation}

Pearson's R ranges between -1 and 1 representing, respectively, (perfect) opposition and similarity. Pearson's R values larger than 0 indicate a positive association between both vectors, whereas values smaller than 0 indicate negative association. As the Pearson's correlation coefficient is only a general measure of the relation between both vectors, a visual investigation of the point cloud of both vectors is used as an additional tool in understanding their relation (see figure \ref{fig:overzicht} A).      

%\subsubsection{Cosine similarity with ground truth data}

%A second way to compare similarities between vectors of the same length is by means of the Cosine Similarity Measure (CSM), which basically derives the angle between two vectors x and y, and expresses it in degrees (°)

%\begin{equation}
%CSM(\bar{x},\bar{y})= \left | cos^{-1} \left (\frac{\bar{x} \cdot \bar{y}}{\left \| \bar{x} \right \| \cdot \left \| \bar{y} \right \|} \right ) *\frac{180}{\pi} \right |
%\end{equation}

%A CSM value of 0° denotes the highest possible similarity between both vectors, 90° indicates the lowest similarity and 180° degrees refers to an opposite orientation.

%\iffalse
%\begin{equation}
%CS(\bar{x},\bar{y})=\frac{\bar{x} \cdot \bar{y}}{\left \| \bar{x} \right \| \cdot \left \| \bar{y} \right \|}
%\end{equation}

%http://www.codecogs.com/latex/eqneditor.php for equation development

%Values of the cosine will range between -1 and 1. A value of 1 indicates the highest similarity in orientation (the angle between x and y is zero degrees), 0 indicates the lowest similarity in orientation (the angle between vector x and vector y is 90 or -90 degrees) and -1 indicates an opposite orientation (the angle between x and y is 180 degrees). Deriving the angle between both vectors and expressing it in degrees (°) results in the more easily interpretable Cosine Similarity Measure (CSM):
%\fi

\subsubsection{LogRatio and spatial patterns (figure \ref{fig:overzicht} B)}
While Pearson's R evaluates similarity between the vectors of population and user counts, it does not entirely allow for an assessment of the differences between ground truth and HDA results for each cell tower. To explore such individual differences, we calculate the ratio of the HDA user count to the ground truth population count, and further take its logarithm because of the right long-tail in the distribution of the ratio. Therefore, the LogRatio for cell tower $i$ is defined by:

\begin{equation}
LogRatio_i =ln(\frac{x_i}{y_i}),
\end{equation}
where $x_i$ is the user count estimated by one HDA, and $y_i$ is the population attributed to cell tower \textit{i} based on census data. Note that, although local market shares of Orange can differ from cell tower to cell tower, LogRatios are distributed around -1.27 (=ln(0.28)), given the 28\% overall market share. Large deviations from the expected 0.28 ratio can therefore be considered indicative for severe under- or overestimations. The LogRatio values for each HDA can then be mapped to directly investigate the spatial pattern of differences between HDAs' user counts and the ground truth population counts, to produce insights on the performance of HDAs in general.

%\subsubsection{Spatial patterns (figure \ref{fig:overzicht}b)}

\subsubsection{Temporality and sensitivity (figures \ref{fig:overzicht} C,D,E)}
To assess the sensitivity of HDAs, we deploy home detection for all combinations of criteria choice, parameter choice and time periods previously defined and investigate performance measures over time (figure \ref{fig:overzicht} C), between different HDAs (figure \ref{fig:overzicht} D) and for different time periods (figure \ref{fig:overzicht} E). As each summarizing measure describes the performance detecting homes for 18 million users, distributed over a total of 18,273 cell towers, small differences in general performance relate to large absolute numbers of users being allocated to a different home location.

\section{Results}\label{sect.results}

\subsection{Relations between HDA's user counts and ground truth}

\begin{figure}[htbp!]
\centering
\caption{Scatterplot of ground truth population counts and user counts based on the MA algorithm for the 14-day period between 25/06 and 08/07. Each dot represents one cell tower, and is colored by density of the dots in its surrounding, in a gradient from blue (low density) to red (high density). For each group of cell towers based on the deciles (except the 90-100 decile) of the population counts (x-axis), the mean (black middle dot) and standard deviation (whisker) of the user counts (y-axis) were calculated and plotted.}
\label{fig:scatterplot}
\resizebox{0.6\textwidth}{!}{ 
\includegraphics{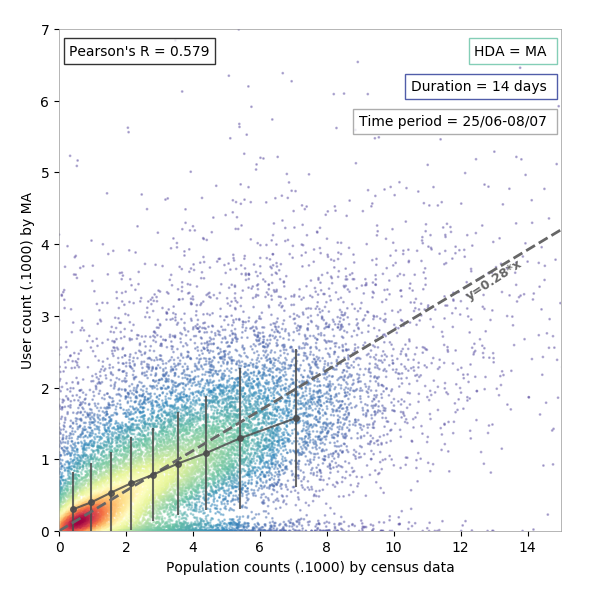}
}
\end{figure}

After running the nine HDAs on all users in the CDR dataset for all time periods, we find the Pearson correlation coefficients defined in Section \ref{sect.pearson_definition} to range between 0.45 and 0.60, which indicates a moderate performance (see also figures \ref{fig:scatterplot},\ref{fig:corr_temporality},\ref{fig:duration_sensitivity},\ref{fig:criteria_sensitivity}). Figure \ref{fig:scatterplot} is exemplary for the relation between user counts obtained from HDAs and the ground truth population counts. While it only depicts the results from the MA algorithm run for the 14-day period between June 25 and July 08, 2007, other HDAs and time periods show similar relations. Three elements in figure \ref{fig:scatterplot} deserve highlighting.

First, the majority of cell towers have rather low population and user counts (between 0 and 2000 persons). This effect is partly due to high densities of cell towers in urban areas. 
Secondly, cell towers are typically centered around the 0.28 line (dashed), which aligns with a 28\% overall market share of the Orange operator, but divergence is high. This is indicated by the means (black dots) and the standard deviations (vertical error bars) of sub-groups of cell-towers grouped according to the deciles of the population counts. \textit{Overestimation} typically occurs for cell towers with lower ground truth population, while \textit{underestimation} is more related to cell towers that have a higher ground truth population. Without more contextual information on the cell towers such as location or typical usage by mobile phone users, it is impossible to properly account for this pattern.
Thirdly, one can clearly distinguish a group of cell towers that have very low user counts. This group seems to be evenly distributed over the different ground truth population counts and can be considered an artifact of data collection. They correspond to cell towers that were (temporally) inactive during certain observation periods because they were, for example, under repair or temporally dislocated.

\subsection{Spatial Patterns of LogRatio}

\begin{figure}[htpb!]
\centering
\caption{Map of the individual LogRatios of all cell towers based on the MA algorithm for the 14-day period between 25/06 and 08/07.}
\label{fig:map_logratio}
\resizebox{0.6\textwidth}{!}{ 
\includegraphics{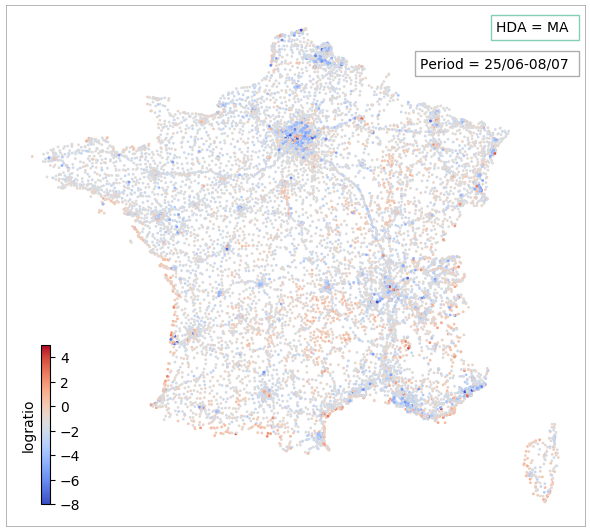}
}
\end{figure}

Figure \ref{fig:map_logratio} shows the spatial pattern of how HDA's user counts overestimate or underestimate population counts, by mapping the LogRatios between both for the MA algorithm deployed on the 14-day period between 25/06 and 08/07 (similar to figure \ref{fig:corr_temporality}). Again, this example is exemplary for other HDAs and time periods. In general, HDAs underestimate populations in major city centers and among major roads compared to ground truth data, while overestimation occurs more in rural areas as well as touristic areas. Overestimation in touristic regions becomes even more prevalent during holiday periods as will be shown in figure \ref{fig:one_hda_two_maps}. Underestimation in city centers can very well relate to local market shares of Orange because city centers are highly competitive locations between operators (with better services offered by smaller operators), resulting in smaller market shares for all operators. Overestimation in rural areas, then again, can be explained by reasons of (historical) coverage and brand loyalty of small communities.

\subsection{Temporality of correlations}

\begin{figure}[htpb!]
\centering
\caption{Performance over time of different HDAs for time periods that have a fixed duration of 14 days. The correlation measures were obtained from comparison with ground truth data (see section \ref{sect.pearson_definition}) and are plotted in the middle of the observation periods, with blue dotted lines and blue arrows indicating the extent of the time period. Other HDAs depict similar temporal patterns.}
\label{fig:corr_temporality}
\resizebox{0.8\textwidth}{!}{ 
\includegraphics{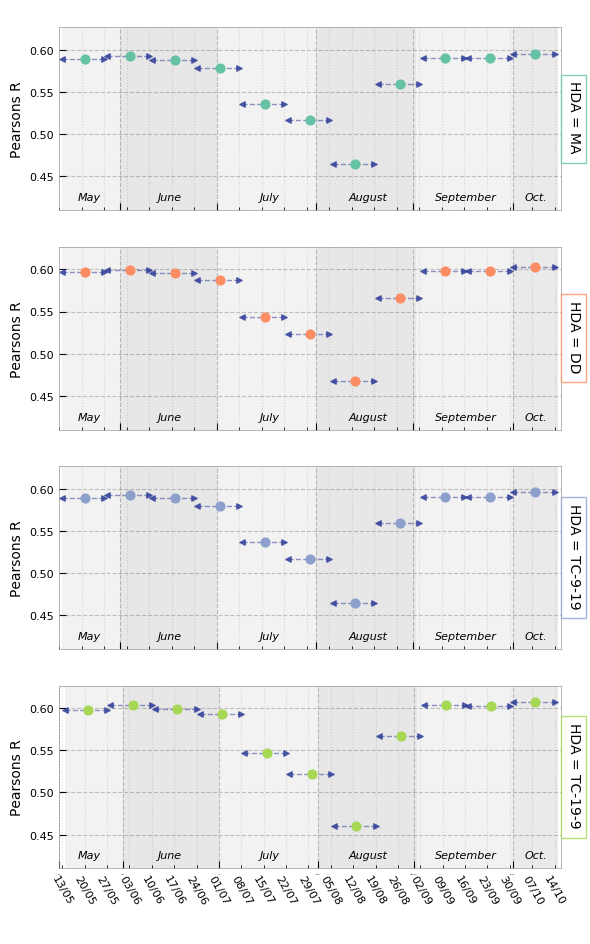}
}
\end{figure}

Investigating Pearson's R values for different HDAs over time, we find the performance of all HDAs to be extremely sensitive to the period of observation. In figure \ref{fig:corr_temporality} that shows the correlation coefficient of four different HDAs for all periods with a duration of 14 days, there is a clear drop in performance during summer months July and August. This drop in performance is observed for all HDAs and is independent of the duration of observation (see also figures \ref{fig:duration_sensitivity} and \ref{fig:criteria_sensitivity}). It is, in magnitude, the largest sensitivity that we observed during our experiments.  

\begin{figure}[htpb!]
\centering
\caption{Spatial patterns of the user counts obtained by the MA algorithm for a non-summer (A) and a summer (B) period of 14 day duration. Home detection in summer periods results in higher user counts in touristic areas, which in turn results in lower correlation with the ground truth dataset(C).}
\label{fig:one_hda_two_maps}
\resizebox{0.8\textwidth}{!}{ 
\includegraphics{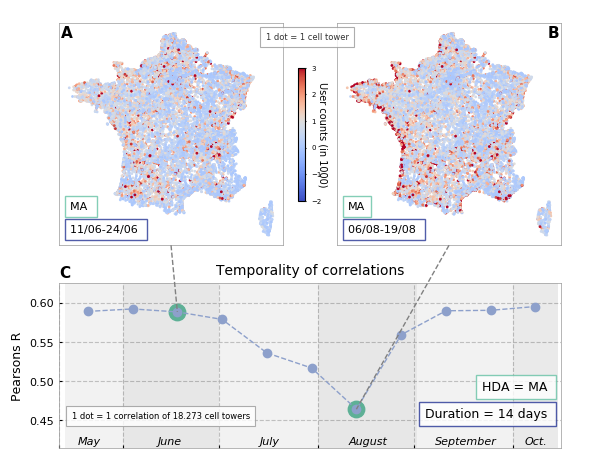}
}
\end{figure}

The limited performance of home detection during summer months is not surprising given that a large share of the French population is known to perform tourism trips to holiday destinations during this period \cite{Deville2014,Vanhoof_domestictourism,Vanhoof_JOS}. It is, however, surprising that this mobility influences the results of HDAs to such a large degree, especially in comparison to the other sensitivities examined. The loss in performance of HDAs during summer periods because of (domestic) tourism \footnote{In 2015, 88.1\% of all tourism trips performed by French people were estimated to be domestic tourism trips; making it one of the largest shares in Europe \cite{Vanhoof_domestictourism}.} can be superbly illustrated by showing the spatial pattern of user counts obtained from a single HDA deployed on both a non-summer and summer period in figure \ref{fig:one_hda_two_maps}. Clearly, during summer periods, user counts resulting from HDA go up drastically in touristic regions such as seaside locations, resulting in a significantly lower correlation with ground truth. 

\subsection{Sensitivity to the duration of observation}

\begin{figure}[htpb!]
\centering
\caption{Performance over time of 4 algorithms for time periods characterized by different durations of observation.}
\label{fig:duration_sensitivity}
\resizebox{0.80\textwidth}{!}{ 
\includegraphics{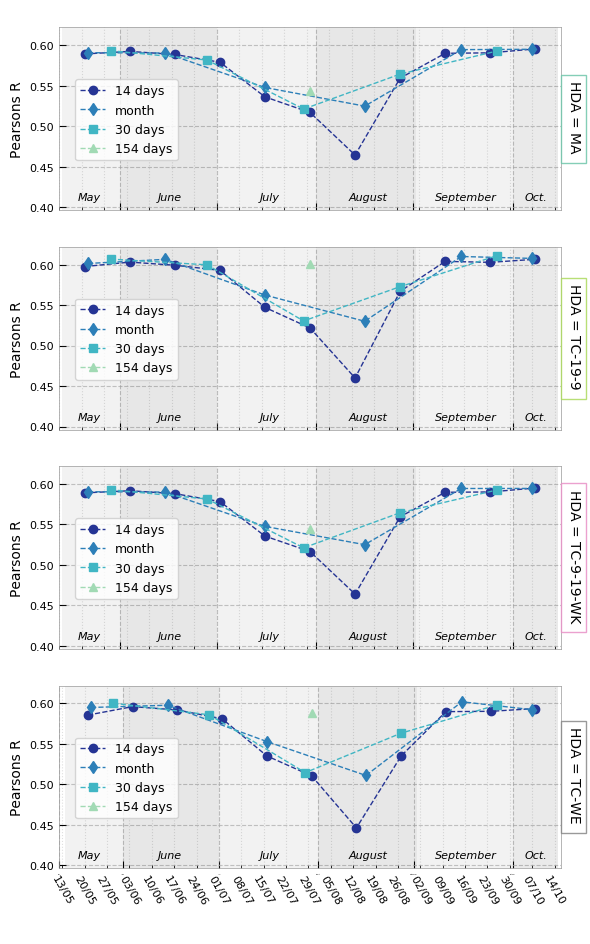}
}
\end{figure}

One outstanding question is to which degree the duration of the time period used is influencing the performance of HDAs. For example, it seems logical that while holiday mobility is influencing home detection, larger durations of observation periods could mitigate for this effect. As can be observed in figure \ref{fig:duration_sensitivity}, which shows the sensitivity of HDA performance to different durations of time periods for 4 different HDAs, longer durations of observation periods indeed do mitigate the summer drop in performance, although only to a certain extent. Interestingly, the sensitivity to the duration of observation is subordinate to the sensitivity to time period (in our case, summer periods), in the sense that duration is influencing performance mainly by the proportion of the time period that occurs in August and, to a lesser degree, in July. 

More specifically, we find that HDAs using shorter durations of observations (such as the 14-day duration) perform amongst the worst when observations are made during summer but amongst the best when the observations are made outside summer months. This is in contrast to the month and 30-day durations, where performance is somewhere in between, depending on the proportion of the time period that is in July or August and independent from the deployed HDA. For the 154-day duration, we find that for some HDAs, the longer time period is capable of mitigating the effect of summer holidays, leading the performances similar to the best performance of shorter durations (for example the TC-19-9 and TC-WE algorithms). This however is not true for all HDAs. Performances for the 154-day duration of the MA and TC-9-19-WK algorithms, for example, are not even close to the performance levels of shorter duration periods that are not occurring in summer. 

The consequence is that there is no clear single duration of observations that performs best, and that all is dependent on the combination of duration, general mobility of the population and the chosen HDA. One rule of thumb could be that, if no insights are available on periods of general mobility, performing home detection on longer durations might be the safest choice, but will probable lead to moderate performance compared to shorter periods of observations of which one can be sure that they are free of large-scale mobility.

\subsection{Sensitivity to criteria choice}

\begin{figure}[htpb!]
\centering
\caption{Performance over time of HDAs with different criteria and/or parameters for all time periods with a 14-day (A,B,C,D) and 154-day (E,F,G,H) duration of observation.}
\label{fig:criteria_sensitivity}
\resizebox{1\textwidth}{!}{
\includegraphics{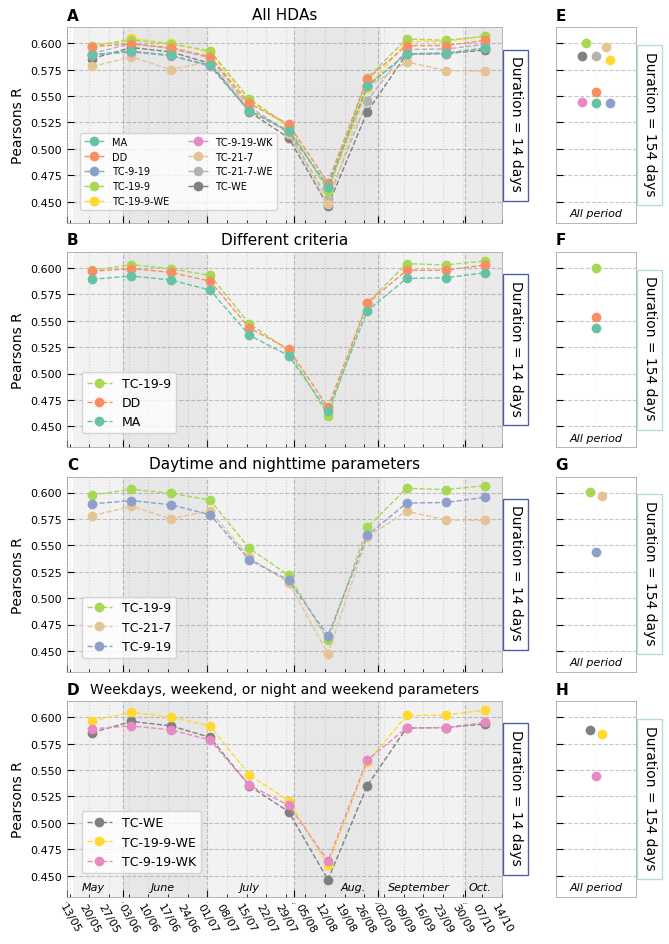}
}
\end{figure}

Final investigation is on the sensitivity of HDA performance to criteria and/or parameter choice. It forms an interesting observation that criteria and parameter choice are less influential compared to time period or, sometimes, even duration of observation choice. 
For periods with a 14-day duration, for example, the effect of criteria choice is about 0.025 (expressed in Pearson's R) whereas the summer period effect is about 0.15, or thus an order higher (figure~\ref{fig:criteria_sensitivity} B). 

Still, some interesting observations can be made when comparing the performance of different HDAs, as was done in figure~\ref{fig:criteria_sensitivity}. One observation is that the TC criterion outperforms the MA and DD criteria for some parameters (such as the 19-9) but definitely not for all. In other words, parameter choice for the TC criterion does have an impact on performance. For example, defining nighttime between 21-7 hours instead of 19-9 hours results in substantial performance loss for all 14-day periods investigated (figure~\ref{fig:criteria_sensitivity} C). Even more remarkable is that the 21-7 parameter, at least for 14 days durations, is consistently outperformed by the 9-19 parameter, which is a daytime definition (figure~\ref{fig:criteria_sensitivity} C). This finding drastically challenges the assumption that using nighttime would be better because people are more at home then. The performance of different TC parameters is also influenced by the time period. Using nighttime and weekends, for example, outperforms using weekdays during non-summer periods, but the true for all 14-day duration periods in August (see figure \ref{fig:criteria_sensitivity} D).

The most remarkable finding regarding the sensitivity of performance to criteria and parameter choice is that it is strongly dependent on the duration of observation. While performance during 30-day and month durations are similar to the 14-days period in figures~\ref{fig:criteria_sensitivity} A,B,C,D), figures~\ref{fig:criteria_sensitivity} E,F,G and H show how patterns of performance change drastically when considering the 154 days duration. Note, for example, how similar the performance of the TC 9-19, TC 19-9, and TC 21-7 are for all 14-days periods (figure~\ref{fig:criteria_sensitivity} C) and how different their performance is for the 154-day period (figure~\ref{fig:criteria_sensitivity} G). 

For the 154 day duration, the general logic seems to be that the TC criterion with parameters restricting to night-time and weekend-days perform significantly better compared to other HDAs (figure~\ref{fig:criteria_sensitivity} E). From our experiments, it is not clear what exactly is driving this gain in performance during longer observation periods, nor is it clear from which duration of observation this gain comes to be (although it has got to be somewhere between 30 and 154 days). One possible reasoning is that TC criteria need an observation period with sufficient duration in order to operate properly according to their semantics, but this requires further inquiry

%\begin{figure}[htpb!]
%\centering
%\caption{Performance over time of HDAs with different criteria for all time periods with a 154-day duration of observation.}
%\label{fig:criteria_sensitivity_154}
%\resizebox{0.80\textwidth}{!}{
%\includegraphics{plot_criteria_sensitivity_moving_154_zoom.png}
%}
%\end{figure}

\section{Discussion and Conclusion}  

Our experiments have revealed the different effects of criteria choice, parameter choice, duration of observation period and time period on the performance of HDAs for a French CDR dataset. We find that HDAs in France perform moderately at best, with correlations with high-level ground truth data ranging between about 0.40 and 0.60. Investigating the spatial pattern of performance, we observe that all nine HDAs for all 225 deployed time periods underestimate populations in city centers and among main roads, while overestimating population in touristic destinations and some rural areas. Our definitions of under- and overestimation do, however, base on an assumed market share of the operator of 0.28\%, which is correct for the national level but can vary locally. Consequently, the observed spatial patterns, especially when defined by the LogRatio, can be influenced by the pattern of variations in local market shares, although to an unknown degree. Unknown local market shares can potentially influence the upper bound of correlations with ground-truth too (\cite{Vanhoof_JOS}). As such, we would suggest future research to incorporate information on local market shares when investigating nation wide performance of HDAs. 

For France, 2007, we find HDA performance to be sensitive, in descending order of magnitude, to the deployed time period, the duration of observation and, to a small degree only, to criteria and parameter choice. The largest sensitivity is to the July and August period, when performance drops significantly for all HDAs. During July and August, HDAs consistently overestimate population counts in touristic regions, suggesting that they are influenced by large-scale holiday movement performed by the French population. The effect of the duration of observation is found to be subordinate to this summer effect, with performance of HDAs being directly related to the proportion of the time period occurring in July and August. Shorter periods which are outside the summer months outperform longer durations for which part of the observations were made in July and/or August. The consequence is that there exist no clear advice on what duration of observation to use. When periods of large-scale population mobility are unknown to the researcher, longer durations might be the safest option. When such periods are known, however, avoiding them at all costs seems the best advice, even if this means giving up part of the dataset.  

Our most remarkable finding is that criteria and parameter choice for HDAs seem of little influence, especially compared to the sensitivity to time period and duration of observation. Regarding criteria we find the `distinct day'-criterion the slightly outperform the `amount of activities'-criterion. The performance of the popular `time constraints'-criterion is highly dependent on the chosen parameters. It is remarkable, for example, that the 19-9 hours time constraint outperformed most algorithms, whereas 21-7 (only a few hours difference) performed the worst of all tested algorithms. Additionally, the performance of HDAs based on the `time constraints'-criterion is rather inconsistent, with intuitive (e.g. nighttime and weekends) and contra-intuitive (e.g. only working days) constraints outperforming each other at different time periods and for different durations of observations. All of this makes one wonder why the `time-contraints'-criterion, although popular in literature (\cite{Vanhoof_JOS}), should ever be opted for; if not for our one observation that, for the 154-day period, time-constraint HDAs with intuitive parameters outperformed all other HDAs, with performance equaling the best performances of other HDAs for other periods and durations in our experiment. 

The main contribution of our work is that it offers an insight in the combined effect of user choices on the performance of home detection when deployed on a French CDR dataset. Our work can help other practitioners to decide on suitable home detection algorithms and time periods of observation. We do, however, strongly urge other researchers to reproduce our experiments on their own datasets, mainly because we found performance in France to be most sensitive to summer periods (which probably corresponds with large-scale holiday movement), and partly because of potential influence of unknown local market shares and mobile phone usage patterns, all of which are location and time specific. Additionally, our work can inform the assessment of uncertainty and error related to home detection methods when performed on CDR data or, to a lesser degree, other large datasets of geo-located traces for individual users. Ultimately, our work can contribute to the wider debate on integrating big data in official statistics in, at least, two ways. First, it serves as a reminder that, despite showing large potential, many big data sources and related methods remain in need of quality assessment. Secondly, it is an example that collaboration between academia, private sector and official statistics offices can be extremely fruitful to overcome barriers experienced by one or multiple parties, even though the actual realization of such collaborations is, due to multiple practicalities, never easy.

%\section*{References}
\bibliographystyle{hapalike}
\bibliography{Chapter_sensitivities_of_home_detection.bib}

\end{document}